# High-performance gate-controlled superconducting switches: large output voltage and reproducibility

L. Ruf[1], E. Scheer[1,†], A. Di Bernardo[1,2,*]

1 *Department of Physics, University of Konstanz, Universitätsstraße 10, 78464 Konstanz, Germany.*
2 *Dipartimento di Fisica "E. R. Caianiello", Università degli Studi di Salerno, via Giovanni Paolo II 132, 84084 Fisciano (SA), Italy.*

†Email: elke.scheer@uni-konstanz.de
*Email: angelo.dibernardo@uni-konstanz.de

**Abstract**
Logic circuits consist of devices that can be controlled between two distinct states. The recent demonstration that a superconducting current flowing in a constriction can be controlled via a gate voltage ($V_G$) – gate-controlled supercurrent (GCS) – can lead to superconducting logic with better performance than existing logics. However, before such logic is developed, high reproducibility in the functioning of GCS devices and optimization of their performance must be achieved. Here, we report an investigation of gated Nb devices showing GCS with unprecedently-high reproducibility. Based on the investigation of a statistically-significant number of devices, we demonstrate that the GCS is independent of the constriction width, in contrast with previous reports, and confirm a strong correlation between the GCS and the leakage current ($I_{leak}$) induced by $V_G$. We also achieve a voltage output in our devices larger than the typical values reported to date by at least one order of magnitude, which is relevant for the future interconnection of devices, and show that $I_{leak}$ can be used as a tool to modulate the operational $V_G$ of devices on a $SiO_2$ substrates. These results altogether represent an important step forward towards the optimization of reproducibility and performance of GCS devices, and the future development of a GCS-based logic.

# I. Introduction

The control of a superconducting current (supercurrent) via the application of a gate voltage ($V_G$), currently known as gate-controlled supercurrent (GCS), has become a subject of great interest, as evidenced by the number of experimental [1-29] and theoretical studies [30-36] reported on it. Amongst the main motivations behind the interest in the GCS is its potential for the development of voltage-controlled superconducting logics that would intrinsically have low-energy dissipation (since based on superconductors) and better performance than other superconducting logics already available [37-39]. Compared to these, GCS-based logics would offer several advantages including higher device density (due to the smaller device size), larger number of devices connectable in series (i.e., higher fan out), stronger resilience to magnetic noise and easier interfacing with conventional metal-oxide semiconductor (CMOS) circuits to form hybrid superconducting/semiconducting computing architectures [40]. Also, superconducting GCS devices can find applications in emerging quantum technologies, and particularly in superconducting quantum processing units (QPUs). As recently suggested [40], GCS devices could be integrated into QPUs and used, for example, as tunable elements to switch on/off the control of multi-qubit gates. In addition, these devices could be integrated also into auxiliary systems (filters, resonators etc.) for the multiplex routing of microwave signals to the QPU.

To develop some of the above applications based on the GCS, however, it is first necessary to overcome fundamental challenges and to meet proper technological standards, both in terms of performance of the individual GCS-based devices and in terms of optimization of their fabrication process.

Some of the current fundamental challenges stem from the lack of a clear understanding of the mechanism responsible for the GCS, which is in turn crucial to achieve control over the effect. As reported in a recent review on the GCS [40], four main mechanisms have been proposed to date to explain the GCS. Some of these mechanisms ascribe the GCS to phenomena triggered by the finite leakage current ($I_\text{leak}$) that flows in most device realizations between the gate and the superconductor (S) constriction under an applied $V_G$. These phenomena include tunneling of high-energy electrons between the gate and the S [12-14,23-24], phonons induced by $I_\text{leak}$ in the substrate heating the S constriction [21], and high-energy electrons or phonons induced by $I_\text{leak}$ which drive the S into an out-of-equilibrium state characterized by phase fluctuations but without heating [19,22]. In addition to these $I_\text{leak}$-related mechanisms, it has also been suggested that the electric field (associated with the applied $V_G$) can induce effects responsible for a GCS [1-10,15,18,25].

It is worth noting that the diversity in the mechanisms proposed to explain the GCS partly originates also from the significant differences in materials (i.e., S type, substrate etc.), device geometry and experimental setups used in the different studies , which makes it difficult also to find a "universal mechanism" at play in all studies on GCS. At this stage, it is therefore crucial to carry out studies where all material and device parameters are kept fixed, and only one parameter at a time is systematically varied, to rule out and/or confirm the main findings made in other reports.

In terms of performance, the voltage $V_{G,offset}$ needed to induce a full suppression of the critical supercurrent ($I_c$) and switch a GCS device to its normal state, must be reduced to few volts, to better interface GCS devices with CMOS devices (typically working at $V_G < 5$ V [41]) in hybrid computing platforms. A lower $V_{G,offset}$ would also lead to an increase in the fan out, since the output voltage $V_{out}$ of a GCS device, which depends on its characteristic voltage $I_c R_N$ ($R_N$ being the normal-state resistance), can be fed as input signal to the gate (i.e., used as the $V_G$) of another GCS device connected downstream [40]. In most GCS devices reported to date, however, $V_{G,offset}$ is typically of few tens of volts [1-2,6,9-10,17,25,27], whilst $I_c R_N$ is typically of few millivolts or tens of millivolts [1-2,4,6,15,20,22,27], meaning that these two voltages differ by several orders of magnitude. This large difference between $I_c R_N$ and $V_{G,offset}$ currently makes the connection of GCS devices in series not feasible.

In terms of the fabrication process, it is necessary to optimize protocols for high scalability of devices and to achieve high reproducibility in their realization of GCS. Several groups have already tried to scale up the fabrication of GCS devices using subtractive patterning [23,26,27], which involves lithographic patterning of a device into a negative resist layer used then as mask to etch an underlying S thin film, other than by additive patterning involving lithographic patterning of a polymer mask followed by S material growth and lift off. Some of these studies, however, suggest that devices made by subtractive patterning do not show a GCS unlike devices made by additive patterning [26], unless unconventional Ss that can be grown with a small grain size (e.g., $Nb_{0.18}Re_{0.82}$) are used, in combination with a non-trivial surface chemistry activated by the etching process [27].

Achieving high reproducibility in the behavior of GCS-based devices, which is also essential for a better fundamental understanding of the effect as explained above, remains another crucial objective. All the studies on the GCS reported to date are in fact based on the characterization of few devices (typically one or two devices in a given study), across which several parameters (e.g., substrate, S type, gate-to-channel distance, device geometry) are often varied, even within the same study [40]. This large variation of parameters makes it difficult to agree on the existence of universal features of the GCS, since certain observations are not reproduced when GCS devices with different parameters are studied.

Here, we address some of the open challenges listed above in the field of GCS. First, we report a fabrication process that allows to systematically reproduce the phenomenon in all our investigated (thirteen) devices. The high reproducibility achieved also allows us, by varying one parameter at a time across different devices, to confirm or demystify sporadic observations, which are often based on a less statistically-relevant number of samples. For example, by progressively varying the width of our S constriction ($w_S$), whilst keeping the other device parameters identical, we show that GCS does not require $w_S$ of the same order of magnitude as the S coherence length for it to be observed, in contrast with previous observations [1,13]. Thanks to an in-depth analysis of other performance parameters of our devices, we also rule out $I_{leak}$-induced Joule heating as possible mechanism for the GCS but confirm the existence of a strong correlation between the suppression of $I_c$ and variations in $I_{leak}$, consistent with other reports [28].

Second, the large $w_S$ used (up to 550 nm) results in devices with $I_c R_N$ up to ~ 0.25 Volts at 1.5 K, which represents a significant step forward towards the increase in the voltage output, as described above. Last, concurrently with an increase in $I_c R_N$, we show that stress induced in the SiO$_2$/Si substrate by $I_{leak}$ can be used as a tool to reduce $V_{G,offset}$. This result does not only possibly explain differences reported in $V_{G,offset}$ even across nominally identical devices and measured under the same conditions [2,22], but it also provides a possible route to explore for the control of the operational $V_{G,offset}$ of a device after its fabrication.

## II. Results and Discussion
### a. Reproducibility of GCS in Nb devices

To investigate the reproducibility of the GCS in three-terminal superconducting devices, we have fabricated a series of gated Nb Dayem bridges (Fig. 1a) on a SiO$_2$ (300 nm)/undoped Si substrate. Across our devices, we have kept all design parameters fixed, except for $w_S$. The parameters that we have kept fixed include the thickness of the S layer (27-nm-thick Nb deposited onto ~ 4 nm of Ti used to promote adhesion), the shape of the gate (pointy), the gate distance from the Nb nanoconstriction (~ 50 nm), and the length of the Nb nanoconstriction (~ 1.2 μm). For $w_S$, we have changed its value from 190 nm to 550 nm across our devices. The list of all the devices investigated in this study with the corresponding geometry and other experimental parameters obtained from their low-temperature characterization is reported in Table S1 of the Supplementary Information.

Fig. 1a shows a scanning electron microscope image of one of the Nb Dayem bridges with a sketch of the four-terminal probe setup used to measure the current versus voltage, $I(V)$, characteristics, whilst applying a $V_G$ to the side-gate electrode. In our setup, the actual gate voltage applied, hereby named $V_G^*$, is not simply equal to the $V_G$ applied by the source meter (see Methods). Since the gate electrode and the negative terminal ($I^-$) of the bias current, $I_{bias}$, are connected to the same electrical ground (Fig. 1a), $V_G$ is shifted from $V_G^*$ by the voltage drop on the wiring resistance ($R_{wiring}$), meaning that $V_G^* = V_G - I_{bias} \cdot R_{wiring}$ (Fig. 1a). Due to the large $I_{bias}$ that we apply in our devices to match $I_c$ (up to 3 mA), the shift in $V_G$ can be as large as 6 V (see Methods). In the following, we therefore show the actual $V_G^*$ applied to the S constriction, after correcting for the voltage drop on $R_{wiring}$ when $I_{bias} = I_c$.

The data reported in Figs. 1b to 1e for a few representative devices with different $w_S$ show that we have achieved a systematic observation of a full superconducting transition and of the GCS in our devices (see also Table S1 and Fig. S1). We have investigated a total of 13 samples with a superconducting critical temperature ($T_c$) ranging from 4.4 K for $w_S$ = 190 nm to 7.57 K for $w_S$ = 550 nm (Fig. 1b and Fig. S1), and they all show a GCS. In this paper, we define $T_c$ as the temperature $T$, where the device resistance $R$ reaches 10% of its normal-state ($R_N$) value at 10 K. This definition yields lower $T_c$ values than those given by more common $T_c$ definitions based on 90%-$R_N$ or 50%-$R_N$ criteria. The reason why we have chosen this definition is related to the appearance of multistep transition in the resistance versus temperature, $R(T)$, characteristics that can be attributed to the lower thickness of the constriction (resulting from the sputtering deposition of Nb) compared to the thickness of the lateral wider areas. We have also made devices where we have cut the S constriction (on the side opposite to the gate) using a focused ion beam (FIB) with Ga$^+$ ions, to assess the role of current-crowding [42-43] effects on the GCS observation (see section c).

Most of the samples have been deliberately made with $w_S$ much larger than the typical $w_S$ (up to ~ 200 nm) reported in previous studies on the GCS [1,3-10,12,20,22, 25,27-28]. This has been done to verify the argument that a GCS is easier to observe as $w_S$ approaches the coherence length $\xi$ of the S. Our measurement results instead show that GCS can also occur in devices with $w_S$ of 550 nm (Fig. 1e), which is ~ 40 times larger than the $\xi$ of Nb in the diffusive regime ($\xi$ < 15 nm; ref. [44]). We have not tested devices with larger $w_S$, which may in principle still show GCS. We also observe that the gate voltage needed to suppress $I_c$ does not increase as $w_S$ gets larger (Figs. 1c to 1e). These observations altogether suggest that whichever the mechanism responsible for the GCS in our devices is, such mechanism does not get suppressed when $w_S$ greatly exceeds $\xi$.

Our results also imply that GCS does not require a weak superconducting constriction with small $I_c$ to occur. This is crucial for future technological applications of GCS-based devices because, although a reduction in the device lateral dimensions is helpful to increase device density, wider constrictions ensure better device stability against prolonged device operation and thermal cycling.

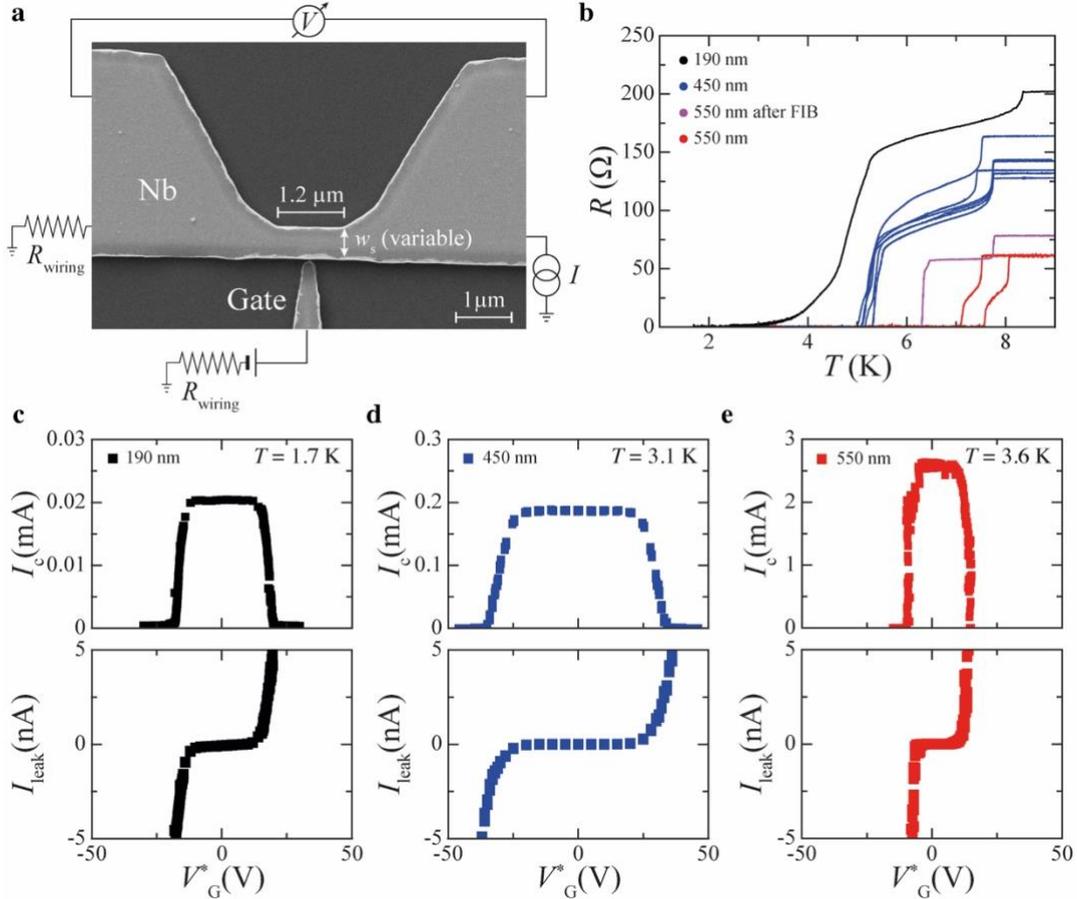

**Figure 1 – GCS in Nb Dayem bridges with different width.** (a) Scanning electron micrograph of a typical gated Nb Dayem bridge with a schematic of the setup used for a four-point measurement of its electronic transport properties. (b) Resistance versus temperature $R(T)$ measured for a few representative Nb Dayem bridges with width $w_S$ of 190 nm (black; 1 device), 450 nm (blue; 6 devices); 550 nm without (red; 2 devices) and with 90-nm-wide cut made by focused ion beam on the S constriction opposite to the gate (purple; 1 device). (c-d) Critical current versus applied effective gate voltage $I_c(V_G^*)$ (top panels) and corresponding leakage current versus $V_G^*$, $I_{leak}(V_G^*)$, curves (bottom panels) measured for a Dayem bridge with $w_S$ equal to 190 nm (c), 450 nm (d), and 550 nm (e). The measurement $T$ is specified in the top-right corner of each panel. $V_G^*$ is defined as $V_G^* = V_G - I_c \cdot R_{wiring}$ where $I_c \cdot R_{wiring}$ is the additional voltage drop induced by a bias current $I_{bias} = I_c$ passing through the wiring resistance $R_{wiring}$ as illustrated in (a).

We note that there is only one report to date [24] about the observation of a GCS in micrometer-wide Nb bridges (i.e., with $w_S \gg \xi$), but these devices have a top-gate other than a side-gate geometry, unlike the ones investigate in the present study. In a device with a top gate, the most relevant dimension for the GCS is not $w_S$ but the S thickness, which in ref. [24] it is ~ 6 nm and hence still comparable to $\xi$ (< 15 nm [44]).

### b. Variations in the GCS with constriction width

The high reproducibility that we have achieved in the observation of a GCS across our devices is not only important as proof-of-principle for future technological applications of the GCS. This reproducibility also allows, by systematically varying only one parameter at a time across different devices, to test whether 1) such parameter is crucial for the GCS observation and 2) if any choices of this parameter improve the device performance (e.g., by reducing the voltage $V^*_{G,\text{offset}}$ needed for a full $I_c$ suppression).

In this study, as also illustrated in Fig. S2, we define $V^*_{G,\text{onset}}$ and $V^*_{G,\text{offset}}$ as the values of $V^*_G$ at which $I_c$ has dropped by 10% and 90%, respectively, compared its value at $V^*_G = 0$ (named $I_{c0}$). All our devices show a $V^*_{G,\text{offset}}$ between ~ 9.5 V and ~ 34 V (Figs. 1c to 1e and Table S1). Only in one device we have measured a much smaller $V^*_{G,\text{offset}}$ (~ 1.6 V), possibly because the $I_{\text{leak}}$ measured at $V^*_{G,\text{offset}}$ for this device is larger by one order of magnitude than that typically measured in the other devices (Table S1).

In Fig. 2, we show the $I(V)$ characteristics measured at a few representative $V^*_G$ for devices with different $w_S$ showing a GCS. Fig. 2c, for example, shows that a device with the largest $w_S$ = 550 nm and hence largest $I_{c0}$ (~ 2.57 mA at $T$ = 3.6 K) has smaller $V^*_{G,\text{offset}}$ compared to a device with $w_S$ = 450 nm and $I_{c0}$ = 0.187 mA at $T$ = 3.1 K (Fig. 2b). In general, we find that the dependence of $V^*_{G,\text{offset}}$ on $w_S$ is not monotonic.

The $I(V)$ curves at different $V^*_G$ in Fig. 2 show that $V^*_{G,\text{offset}}$ increases when going from $w_S$ = 190 nm (Fig. 2a) to $w_S$ = 450 nm (Fig. 2b), and then decreases again from $w_S$ = 450 nm (Fig. 2b) to $w_S$ = 550 nm (Fig. 2c). Also, if a device with original width $w_S$ = 550 nm is reduced to $w_S$ ~ 460 nm by FIB cutting, its $V^*_{G,\text{offset}}$ can even increase (Fig. 2d) compared to the $V^*_{G,\text{offset}}$ measured for a twin device (i.e., made in the same batch) with $w_S$ = 550 nm that has not been FIB-cut (Fig. 2c).

Below in section f, we will show that the presence of conducting paths in the SiO$_2$/Si substrate can induce variations in $V^*_{G,\text{offset}}$. Since the location of these conducting paths cannot be controlled or determined *a priori* and their formation also depends on the device measurement history, these effects can account for the apparent serendipity observed across our devices in their $V^*_{G,\text{offset}}$ dependence on $w_S$.

### c. Independence of the GCS on current crowding

In devices with sharp edges between the S constriction and the lateral pads, current-crowding effects may appear [42-43], which can in turn affect the GCS. In our Nb Dayem bridges, we have tried to minimize current-crowding effects by deliberately fabricating devices with rounded corners at the intersection between the S constriction and the lateral pads, (Fig. 1a). The term current-crowding denotes the observation that sharp edges and sudden changes of the cross section result in a locally-enhanced current density, which can in turn locally overcome the critical current density of S.

To investigate whether current crowding plays any roles in the GCS of our devices, we have also deliberately introduced a sharp edge in one of our devices around which current-crowding effects should become more prominent. This sharp edge has been made by cutting out part of the S constriction using FIB, after the deposition and lift-off of Nb (see Methods). As shown in Fig. S3, the FIB cut (~ 90 nm in width and ~ 80 nm in length) has been made on a device originally with $w_S$ = 550 nm on the constriction side opposite to the gate. This has been done to minimize Ga$^+$ implantation in the channel between gate and constriction, which may lead to an increase in the $I_{leak}$ and in turn affect the measured $V^*_{G,offset}$.

The measurement data in Fig. 2c and Fig. S3 show that, although $I_{c0}$ is reduced for the FIB-cut device by almost 50% compared to the twin uncut device (due to the $w_S$ reduction and possible FIB-induced damage), $V^*_{G,onset}$ and $V^*_{G,offset}$ have increased by more than 250% after the FIB cut (see devices D10 and D12 in Table S1). This result suggests that current crowding does not foster or support the GCS.

### d. Large output voltage and other figures of merit of Nb Dayem bridges

As described in the introduction section, another challenge for technological applications of the GCS is to increase the output voltage $V_{out}$ of a GCS device when this is switched to the normal state by an applied $V^*_{G,offset}$, so that $V_{out} > V^*_{G,offset}$ also for a second device connected downstream. In addition to decreasing $V^*_{G,offset}$, another way to meet this condition is through increasing $V_{out}$, which in turn depends on the characteristic voltage $I_{c0}R_N$ of a GCS device.

The $I_{c0}R_N$ values of the GCS devices reported to date are typically in the range from few millivolts to tens of millivolts, as reported in a recent review on GCS [40]. Thanks to the larger $w_S$ compared to those investigated before, our devices also exhibit larger $I_{c0}R_N$ values. Fig. 3a shows that we have obtained $I_{c0}R_N$ larger than 0.15 V for devices with $w_S$ = 550 nm. We have measured $I_{c0}R_N$ ~ 0.16 V at a $T$ = 3.6 K, which is half the $T_c$ of the device (~ 7.1 K), and $I_{c0}R_N$ ~ 0.25 V when the same device (device D11) was cooled down to ~ 1.5 K.

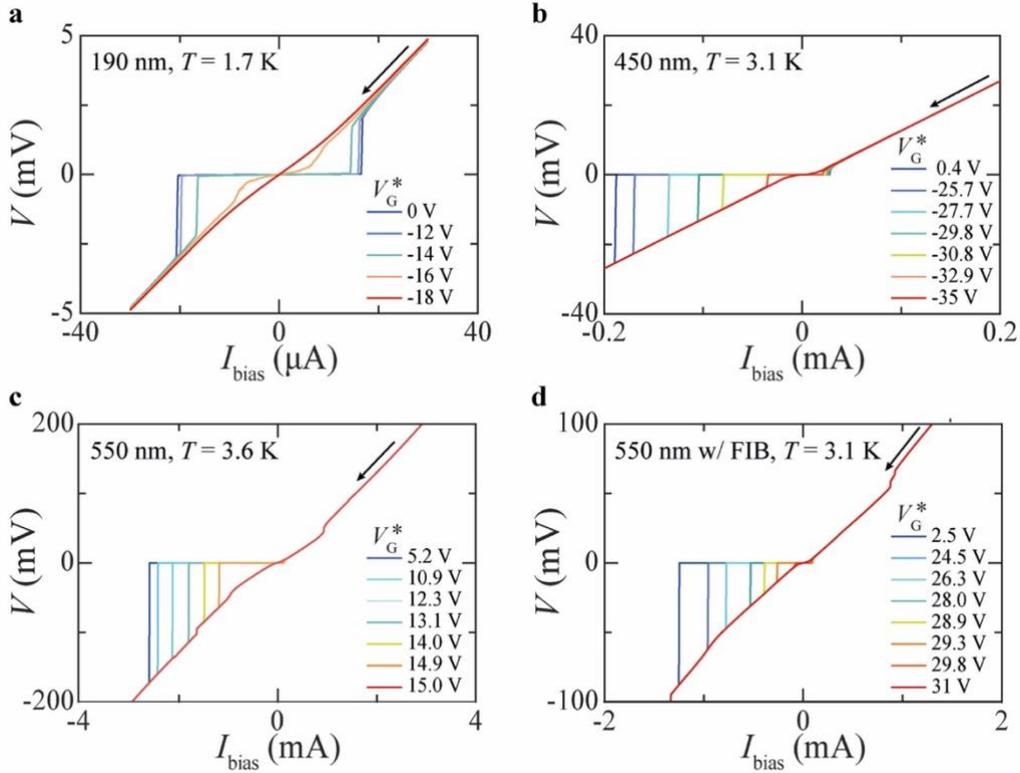

**Figure 2 – GCS for different device widths.** (a)-(d) Current versus voltage, $I(V)$, characteristics as a function of the applied effective gate voltage $V_G^*$ for different S constriction widths $w_S$: 190 nm (a), 450 nm (b), 550 nm (c) and 550 nm after 90-nm-wide cut made by FIB opposite to the gate electrode. All the $I(V)$ curves shown are measured whilst decreasing $I_{bias}$ as indicated by the black arrow in each panel. In each panel, the $V_G^*$ values are defined at $I_{bias} = I_c$ as described in section a, and the measurement temperature is specified in the top-left corner.

Considering that no indication of suppression or saturation of the GCS was observed up to the maximum $w_S > 550$ nm investigated here, we believe that $I_{c0}R_N$ values up to 1 V are within reach, e.g. when using even wider or longer constrictions. Also, even keeping the same device geometry used in this study but replacing Nb with another S like NbN or NbRe with higher normal-state resistivity $\rho_N$ and/or higher critical current density $J_c$, can help increase $I_{c0}R_N$ up to or above 1 V, especially at measurement $T$ smaller than ours.

We now discuss the variation of different figures of merit across our devices with different $w_S$. The data in Fig. 3a and Table S1 suggest that both $V_{G,onset}^*$ and $V_{G,offset}^*$, to which we collectively refer as suppression voltages ($V_{G,supp.}^*$), show a drop at highest $I_{c0}R_N \sim 0.15$ V obtained for $w_S = 550$ nm. Fig. 3a, however, also shows that the lowest $V_{G,offset}^*$ is obtained for a device with the smallest $w_S = 190$ nm fabricated in our study. Therefore, we conclude that no systematic dependence of $V_{G,supp.}^*$ on $I_{c0}R_N$ can be inferred.

A more systematic trend can be instead observed in the dependence of the power dissipated by the gate at $V_{G,offset}^*$, namely $P_{G,offset}^* = V_{G,offset}^* I_{leak,offset}$ (with $I_{leak,offset}$ being $I_{leak}$ measured at $V_G^* = V_{G,offset}^*$), on $I_{c0}R_N$. Fig. 3b in fact shows that $P_{G,offset}^*$ increases almost linearly with $I_{c0}R_N$, which gets in turn larger for increasing $w_S$. We therefore conclude that a larger $P^*_{G,\text{offset}}$ is needed to trigger the GCS in wider devices. This result *per se* can suggest that power dissipation, which scales with $I_{\text{leak}}$, can play a non-negligible role towards the GCS in our devices. We note that the data in Figs. 3b to 3d refer to devices that have all been measured in the same cryostat, so that contributions from the setup to $I_{\text{leak}}$ at a given $V^*_G$ (e.g., due to the shielding of the wires) and in turn to $P^*_G$ are constant. The same data of Figs. 3b to 3d but for all devices investigated in this study (i.e., including those measured in different cryostats) are reported in Fig. S4.

Based on the significant number of devices tested, we have also explored whether there exists a minimum threshold for $P^*_{G,\text{offset}}$, above which a device always shows a GCS independently on $I_{c0}R_N$. To this purpose, we have calculated the ratio between $P^*_{G,\text{offset}}$ and the power dissipated by a device when this switches to the resistive state. For the latter quantity, we have considered both $P_{c0} = R_N I_{c0}^2$ (Fig. 3c) and $P_{r0} = R_N I_{r0}^2$ (Fig. 3d), where $I_{r0}$ is the retrapping current measured at $V^*_G = 0$. We consider both $P_{c0}$ and $P_{r0}$ because the $I(V)$ characteristics of our devices are strongly hysteretic, as shown in Fig. 2, and therefore a device is in a metastable state for $I_{\text{bias}}$ values between $I_{c0}$ and $I_{r0}$. The data in Fig. 2 also suggest that the difference $|I_{c0}| - |I_{r0}|$, and hence in turn $P_{c0} - P_{r0}$, gets larger as $w_S$ is increased.

Figs. 3c and 3d show that, whilst $P^*_{G,\text{offset}}/P_{c0}$ varies across our devices by more than one order of magnitude (from 6 x 10$^{-4}$ for $w_S$ = 550 nm to 9 x 10$^{-3}$ for $w_S$ = 450 nm), $P^*_{G,\text{offset}}/P_{r0}$ shows a small deviation about an average value of ~ 0.31, which seems independent on both $V^*_{G,\text{offset}}$ and $w_S$. This finding is consistent with other reports [13] that have shown that $P^*_{G,\text{offset}}/P_{r0}$ is very similar across different devices, although these studies do not consider devices with large variations in $w_S$ as we do instead here.

The data in Fig. 3c also suggest that, as $I_{\text{leak}}$ increases, the device switches to the normal state when $P^*_{G,\text{offset}}/P_{r0}$ overcomes a specific threshold. Future studies can help understand whether this $P^*_{G,\text{offset}}/P_{r0}$ threshold is set by specific physical or structural properties of the S used or, for example, by the type of substrate on which the device is made (SiO$_2$/Si in all our devices), since the substrate also affects $I_{\text{leak}}$ and in turn $P^*_{G,\text{offset}}$.

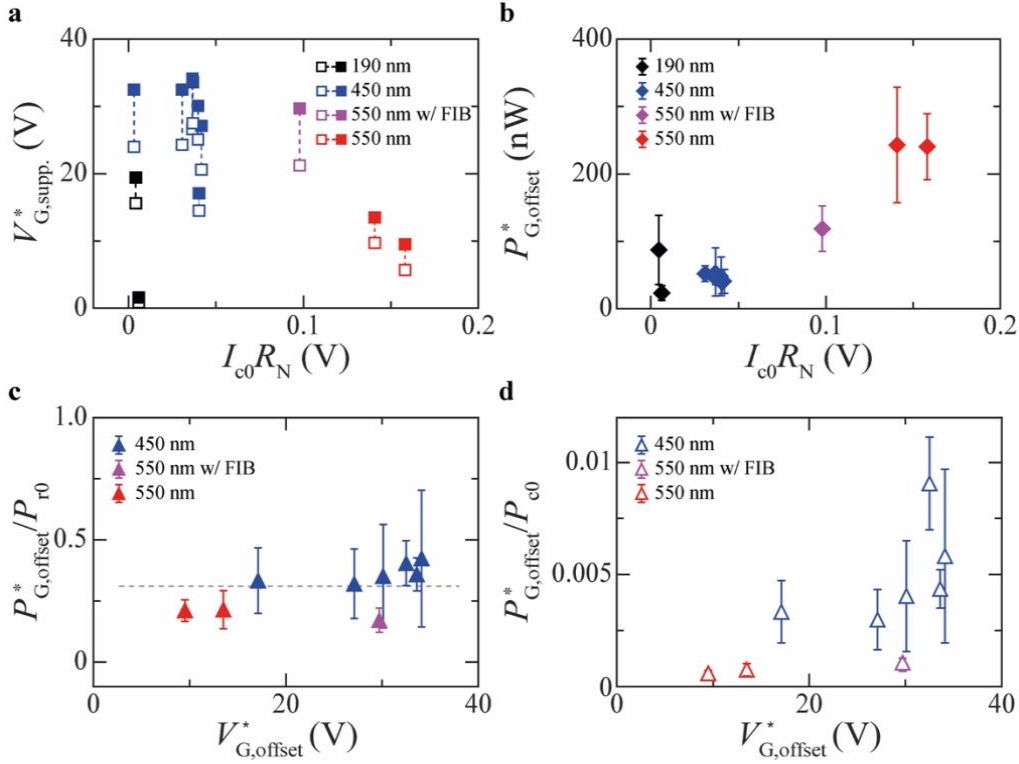

**Figure 3 – Figures of merit of Nb Dayem bridges.** (a) Gate voltage needed to suppress $I_c$, $V^*_{G,supp.}$, at 10% of its value $I_{c0}$ at $V^*_G = 0$ ($V^*_{G,onset}$; empty squares) and at 90% of $I_{c0}$ ($V^*_{G,offset}$; full squares) as a function of $I_{c0}R_N$ for devices with different widths $w_S$ (specified in the legend). (b) Power dissipation induced by the gate voltage at $V^*_{G,offset}$, $P^*_{G,offset}$, as a function of $I_{c0}R_N$ for different $w_S$ values (specified in the legend). (c-d) Ratio between $P^*_{G,offset}$ and $P_r = R_N I^2_{r0}$ (c) and between $P^*_{G,offset}$ and $P_c = R_N I^2_{c0}$ (d) ($I_{r0}$ being the retrapping current at $V^*_G = 0$) as a function of $V^*_{G,offset}$ for devices with different $w_S$ values (specified in the legend). The data points in panels from (b) to (d) refer to devices all measured in the same setup.

The small values of $P^*_{G,offset}/P_{c0}$ measured (Fig. 3c) also suggest that, unlike what one could simply conclude based on the increase of $P^*_{G,offset}$ with $I_{c0}R_N$ (Fig. 3b), power dissipation in the form of simple Joule heating is unlikely the mechanism responsible for the GCS in this study.

### e. Anticorrelation between $I_{leak}$ and $I_c$

To gain further insights on how $I_{leak}$ affects the GCS in our devices, we have also studied if there is any systematic correlations between $I_{leak}$ and $I_c$, meaning whether these two parameters vary independently or not.

To address this question, we have measured the average $I_c$, $I_{c,avg}$, and average $I_{leak}$, $I_{leak,avg}$, from a statistically-significant number of measurements done at each applied $V^*_G$ and calculated the standard deviations of their populations which we name $\sigma_{Ic}$ and $\sigma_{Ileak}$, respectively. Fig. 4a shows that, for $V^*_G > V^*_{G,onset}$, $I_{leak,avg}$ increases (red line), whilst $I_{c,avg}$ decreases (blue line). On the other hand, $\sigma_{Ic}$ also increases, meaning that the switching current distribution (SCD) gets wider due to GCS, as already reported [6,17,19], and this trend is similar for $\sigma_{Ileak}$. Only for $V_G^*$ approaching $V_{G,offset}^*$, $\sigma_{Ic}$ gets reduced [6,17,19] (see also Fig. S5). Fig. S6 shows that also the skewness of the SCD $I_{leak}$ distributions vary in opposite ways for $V_G^* > V_{G,onset}^*$.

In addition to showing opposite trends in the average amplitudes and skewness of their distributions (Fig. 4a and Fig. S6), the change in $I_{leak}$ and $I_c$ also happens concurrently and over long timescales, as shown by time-dependent evolution of their traces reported in Fig. 4b at a fixed $V_G^* \sim 31.9$ V $> V_{G,onset}^*$ (and in Fig. S7 for additional $V_G^*$ values).

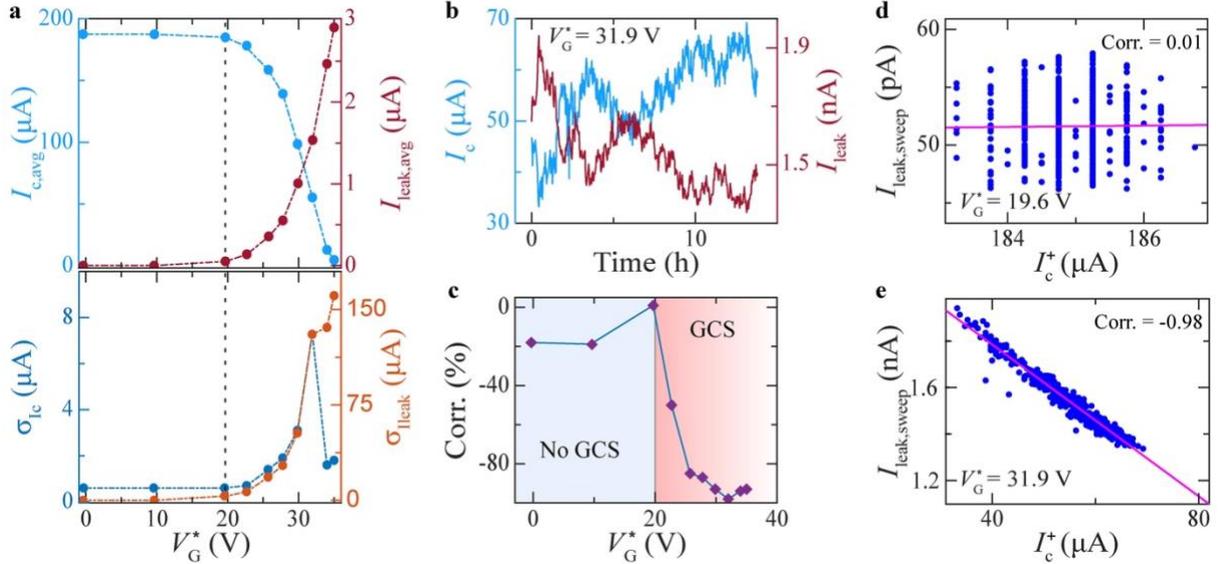

**Figure 4 – Anticorrelation between $I_{leak}$ and $I_c$.** (a) Average critical current $I_{c,avg}$ (top panel, left axis) and average leakage current $I_{leak,avg}$ and standard deviations of the corresponding $I_c$ distribution, $\sigma_{Ic}$, (bottom panel, left axis) and $I_{leak}$ distribution, $\sigma_{Ileak}$, (bottom panel, right axis) as a function of the applied $V_G^*$. (b) Time evolution of $I_c$ and $I_{leak}$ at fixed applied $V_G^* \sim 31.9$ V. (c-e) Correlation factor of average $I_{leak}$ per sweep, $I_{leak,sweep}$, and positive $I_c$, $I_c^+$, as a function of $V_G^*$ (c) and at specific values of $V_G^* = 19.6$ V (d) and of $V_G^* \sim 32$ V (e), respectively, above and below $V_{G,onset}^*$.

The above observations therefore suggest that the amplitudes of $I_{leak}$ and $I_c$ vary simultaneously but in opposite ways, meaning that they are anticorrelated. To get a more quantitative estimate of the anticorrelation between the amplitudes of $I_{leak}$ and $I_c$, we have also determined, at a fixed $V_G^*$, the average $I_{leak}$ measured during a single $I(V)$ upsweep, $I_{leak,sweep}$, as a function of the $I_c$ extracted from the same $I(V)$ curve and repeated this process for several $I(V)$ sweeps. The data obtained are shown for two representative values of $V_G^*$ in Figs. 4d and 4e. At $V_G^* = 19.6$ V, i.e. well below $V_{G,onset}^* \sim 24.3$ V, there is a negligible correlation $\sim 1\%$ between $I_{leak,sweep}$ and $I_c$ (Fig. 4d). However, when a $V_G^* \sim 31.9$ V $> V_{G,onset}^*$ is applied, the anticorrelation between $I_{leak,sweep}$ and $I_c$ reaches a value of 98% (Fig. 4e). The data in Fig. 4c show that this anticorrelation sets in exactly when the GCS also arises at $V_G^* > V_{G,onset}^*$ (see also Fig. S8). The strong anticorrelation that we observe between $I_c$ and $I_{leak,sweep}$ suggests that the GCS in our Nb devices is driven by $I_{leak}$.

**f. Effect of the SiO$_2$ substrate and training of devices**

In addition to a strong anticorrelation between $I_{leak}$ and $I_c$ (at a given applied $V_G^*$), our devices also show strong fluctuations in both $I_{leak}$ and $I_c$ both on a short timescale (i.e., over periods of few seconds) and on a long timescale (i.e., over several hours). We attribute the fluctuations in $I_{leak}$ and $I_c$ to electromigration and/or diffusion processes of atomic species that occur in the SiO$_2$/Si substrate under an applied $V_G^*$. It is well-established that these processes in dielectrics like SiO$_2$ can lead to the formation of metallic weak links that act as paths of low resistance for $I_{leak}$ thus reducing the breakdown voltage of the dielectric [45-46].

The same effects, however, can be achieved not only by applying a $V_G^*$, but also by injecting a high current through the dielectric. This process known as stress-induced leakage current (SILC) has been studied for metal-oxide-semiconductor field-effect transistors down to low temperature, where it has been reported that SILC manifests through fluctuations in $I_{leak}$ occurring over periods of few seconds [47].

In addition to SILC, current-induced stress can also lead to instabilities in $I_{leak}$ associated with fluctuations of single defects under the applied $V_G^*$. This second effect, which can reversibly switch on/off a significant portion of the $I_{leak}$ flowing through a dielectric, is also known as variable stress-induced leakage current (V-SILC) [48-49]. The main difference between SILC and V-SILC is that V-SILC can only induce an additional variation in $I_{leak}$, on top of that induced by SILC.

To understand whether the fluctuations that we observe in $I_{leak}$ at a fixed $V_G^*$ are due to SILC or V-SILC, we have performed a test, where we have probed the time evolution of $I_{leak}$, $I_{leak}(t)$, at fixed $V_G^* = 27$ V, whilst disconnecting all the instruments used for the measurement of the $I(V)$ characteristics. In this configuration, where no $I_{bias}$ is injected, we can assume that no current-induced stress is present, meaning that additional SILC-induced contributions to $I_{leak}$ are negligible. Since we still observe multiple instabilities in $I_{leak}(t)$ other than the bistability expected for a single defect, we infer that the fluctuations in $I_{leak}$ originate from an ensemble of switchable defects in the SiO$_2$. This is also confirmed by the fact that the power spectral density of $I_{leak}(t)$ nearly follows the trend of $1/f^2$-type noise ($f$ being the frequency) suggesting that the $I_{leak}$ values follow a Poisson distribution. This analysis reported in Fig. S9 therefore suggests that $I_{leak}$ most originates from V-SILC due to multilevel switching of several defects in the SiO$_2$ layer (between the gate and the S constriction) under the applied $V_G^*$. We note that similar observations, although on different timescales, have been made for GCS devices made of semiconducting nanowires covered with Al (S) but fabricated also on SiO2/Si substrate like our devices [28]. We therefore attribute the SILC and V-SILC observations to the substrate rather than to the S devices.

Having seen that the fluctuations in $I_{leak}(t)$ at fixed $V_G^*$ are due to V-SILC, we have also deliberately induced SILC in the SiO2 to study if and to which extent it affects the performance of our devices. To induce a SILC effect, on another device made on a fresh SiO2 substrate, we have sourced an $I_{leak}$ between the gate electrode and the S constriction and monitored $V_G^*$, whilst progressively increasing $I_{leak}$. Our idea here is that, as $I_{leak}$ is increased, SILC should manifest and trigger a sudden change in the SiO2 substrate due to formation of weak links, which would in turn lead to a sudden drop of the resistance (and hence $V_G^*$) measured between the gate and S constriction.

Following this protocol, we have observed that a SILC-induced drop in $V_G^*$ indeed occurs as $I_{leak}$ is increased. Once this drop is measured, we then characterize GCS in this new state of the device by measuring a full set of $I_{leak}(V_G^*)$ and $I_c(V_G^*)$ curves. The results of our measurements reported in Fig. 5a show that the $I_c(V_G^*)$ curves measured after a SILC-related change in the SiO2 substrate shift towards a lower $V_{G,offset}^*$, meaning that the GCS sets in at a lower $V_G^*$ value.

Our results therefore suggest that SILC can be exploited in GCS devices as a resource to control their $V_{G,offset}^*$. Using this strategy, meaning by pre-training the devices through the injection of an $I_{leak}$ between gate and constriction, we have achieved remarkable shifts in $V_{G,offset}^*$, $\Delta V_{G,offset}^*$, up to ~ 20 Volts (Fig. 5a). This result also suggests that SiO2/Si is not a good substrate for the realization of GCS devices that would always work with high reliability, meaning in the same operational conditions, because SILC-related effects can occur over time and shift the working point of the GCS device. A fundamental difference between SILC and V-SILC, however, is that SILC shift the working point of the device to a new stable condition, when they occur, whilst V-SILC can give instabilities over short timescales, even after a new SILC-induced working point has been reached.

Although SILC results in a shift of $V_{G,offset}^*$, we find that the power suppression of $I_c$ follows the same exact dependence on the power dissipated by the gate $P_G^* = V_G^* \cdot I_{leak}(V_G^*)$ after each SILC event (Fig. 5c). This is because, although $V_G^*$ decreases after a SILC event, $I_{leak}(V_G^*)$ increases (Fig. 5b), which makes their product constant. Our observation therefore suggests that $P_G^*$ must always reach the same value for the GCS to occur, independently of the history of the device and previous SILC events induced therein.

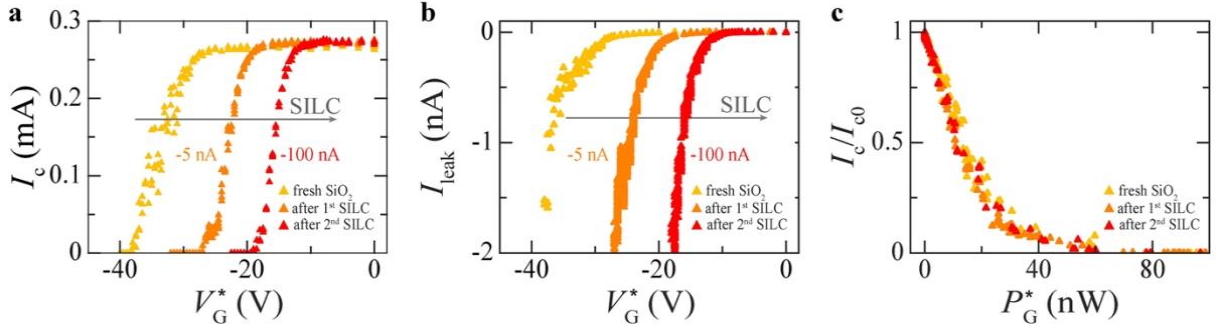

**Figure 5 – Effect of SILC on device performance.** (a-b) Critical current (a) and leakage current (b) as a function of applied gate voltage, $I_c(V_G^*)$, measured after inducing subsequent SILC events through current injection between gate and S constriction. The values of the injected current triggering a SILC event are specified next to the corresponding curves. (c) Dependence of $I_c$ (normalized to its value $I_{c0}$ at $V_G^* = 0$) on the power dissipated by the gate $P_G^* = V_G^* \cdot I_{leak}(V_G^*)$ after each SILC event.

## III. Conclusions

We have shown that three-terminal superconducting devices can exhibit GCS with very high reproducibility, which is a key result for the future development of any technological applications based on the GCS.

Starting from devices that systematically show a GCS, we also find that the effect is not limited by the width $w_S$ of the gated superconducting constriction. Due to their large $w_S$, our devices show unprecedently high $I_cR_N$ values (up to ~ 0.25 V at 1.5 K), which suggests that values above 1V are totally within reach at lower temperatures, especially if devices with a longer and wider constriction and/or made of a S with higher $J_c$ or $\rho_N$ are fabricated. The feasibility of high $I_cR_N$ values shown by our study, together with a reduction in $V_{G,offset}$, represents a significant milestone towards the interconnection of GCS devices and increase in their fan out.

Our analysis also shows that, although $I_{leak}$-induced Joule heating seems to have a minimal effect on the GCS in our devices, the strong correlation between the suppression in $I_c$ (i.e., GCS) and $I_{leak}$ suggests that the GCS is nevertheless $I_{leak}$-induced. Although this result does not allow *per se* to draw definite conclusions on the mechanism responsible for the GCS, we show that, for devices made on SiO$_2$, the strong influence of $I_{leak}$ on the GCS can be exploited to modulate the operational $V_G$ range of the devices. By progressively increasing $I_{leak}$, the devices can be driven through several metastable transitions due to electromigration effects occurring in SiO$_2$ (SILC), which reduces their $V_{G,offset}$ by up to 20 Volts. Independently on the working point of the device and how it is shifted by previous SILC events, however, we find that the $I_c$ suppression always occurs when the power dissipated by the gate voltage (or its relative ratio to the $P_{r0}$) overcomes a specific threshold value. Although a pre-training with SILC events represents a route to modulate $V_{G,\text{offset}}$ for devices made on SiO$_2$, it also suggests that, for technological applications requiring devices with stable operational conditions, substrates different from SiO$_2$ should be used.

**Materials and methods**

**Sample fabrication.** Nb devices with a Ti adhesion layer have been deposited onto a 350-μm -thick (100)-oriented intrinsic Si substrate with a 300-nm-thick wet/dry/wet $SiO_2$ layer on top (MicroChemicals manufacturer). Before growth of the thin films, the substrates (diced into 5x5 $mm^2$ pieces) have been cleaned for 5 min with both acetone and isopropanole (IPA) and then blown-dry with pure $N_2$.

A single 225-nm-thick layer of polymethyl methacrylate (PMMA) (950 PMMA A4, Kayaku) has been then spun onto then substrates and then baked on a hot plate at 180°C for 90 s. After this step, the device geometry (Dayem bridge) has been patterned into the PMMA in a single-step electron beam lithography (EBL) process. The EBL patterning has been carried out using an acceleration voltage of 20 kV and a dose ranging between 280 and 300 μC/cm². Right after exposure, the positive resist mask has been developed by dipping the samples into a methylisobutylketon (MIBK) solution (3 parts of IPA mixed with 1 part of MIBK) for 25 s.

The as-patterned samples have been then loaded into an ultra-high vacuum (UHV) chamber with base pressure lower than $2·10^{-8}$ Torr, where the Ti adhesion layer has been sputtered by radiofrequency (RF) magnetron sputtering. For the Ti deposition, a magnetron gun power of 200 W, an Ar flow of 17 sccm and a deposition pressure of 1.5 mTorr have been used. After the Ti growth, the Nb layer has been sputtered at 300 W (17 sccm Ar flow, deposition pressure of 1.5 mTorr) using two RF magnetron guns and one DC magnetron gun simultaneously. The deposition rates both for Ti (0.055 nm/s) and Nb (0.4 nm/s) have been calibrated using atomic force microscopy.

After deposition, the devices have been immediately placed into a 50 °C hot acetone bath for lift-off for at least 3 hours followed by ~60 s of ultrasonication. Afterwards the devices have been cleaned with IPA and then dried with $N_2$. Before bonding, the Nb devices have been kept under $N_2$ atmosphere to avoid oxidation.

**Transport measurements.** Current versus voltage, $I(V)$, characteristics have been measured with a standard 4-point configuration using a low-noise DC current source, Keithley 6221, to inject the bias current, and a nanovoltmeter, Keithley 2182A, to measure the voltage drop across the Dayem bridge. All Dayem bridges except the ones with $w_S$ = 190 nm have been measured in a dry inverted cryostat (Dry ICE 3K INV) with a base temperature of around 3 K. The measurement lines of this cryostat have been filtered using two-stage RC filters with a series resistance of ~2.05 kΩ and a capacitance of 4 nF.

The devices with $w_S = 190$ nm have been measured in another dry cryostat (Cryogenic Ltd. manufacturer) with a base temperature of ~ 1.5 K reachable with a He$^4$ dip stick and of ~ 300 mK reachable with a He$^3$ dip stick. The lines of this cryostat have been partially filtered using a RC filter with a series resistance of 100 Ω and a capacitance of 47 nF.

The measurement temperature has been chosen to ensure a bath temperature stability of ± 10 mK or better. For wires with 550 nm made with and without FIB, the bath temperature have been increased from 3.1 K (used for other devices) to 3.6 – 4 K (see devices D11 and D12 in Table S1) to ensure enough cooling power and to ensure thermal stability of the measurement system, whilst injecting a large bias current (up to 2.6 mA).

For all devices, the leakage current has been measured using a low-noise source-measure unit, Keithley 6430, with a pre-amplifier connected in a two-wire configuration. For the Dry ICE 3K INV system four extra lines have been used to apply the gate voltage, which are unfiltered and shielded to measure low-leakage currents (at 70 V, these lines have more than 10 TΩ of resistance to ground). For the Cryogenic setup, the measurement lines are made of manganin twisted pairs with a double isolation of Kapton (the lines have more than 50 GΩ of resistance to ground at $V_G > 50$ V).

**Acknowledgements**

We thank Karl Berggren for scientific discussion and Matthias Hagner, Annika Zuschlag, and the staff of the nano.lab at the University of Konstanz for technical support. We also acknowledge funding from the EU's Horizon 2020 research and innovation programme under Grant Agreement No. 964398 (SUPERGATE) and from the University of Konstanz via a Zukunftskolleg research fellowship. A.D.B. also acknowledges support from the MAECI through the grant 'ULTRAQMAT'.

**Competing interests**

The authors declare no competing interests.

**References**

1. G. De Simoni, F. Paolucci, P. Solinas, E. Strambini, F. Giazotto, "Metallic supercurrent field-effect transistor", *Nat. Nanotechnol*. **13**, 802-805 (2018) https://doi.org/10.1038/s41565-018-0190-3
2. F. Paolucci, G. De Simoni, P. Solinas, E. Strambini, N. Ligato, P. Virtanen, A. Braggio, F. Giazotto, "Magnetotransport experiments on fully metallic superconducting Dayem-bridge field-effect transistors", *Phys. Rev. Appl*. **11**, 024061 (2019). https://doi.org/10.1103/PhysRevApplied.11.024061
3. G. De Simoni, F. Paolucci, C. Puglia, F. Giazotto, "Josephson field-effect transistor based on all-metallic Al/Cu/Al proximity nanojunctions", *ACS Nano* **13**, 7871-7876 (2019). https://doi.org/10.1021/acsnano.9b02209
4. F. Paolucci, F. Vischi, G. De Simoni, C. Guarcello, P. Solinas, F. Giazotto, "Field-effect controllable metallic Josephson interferometer", *Nano Lett*. **19**, 6263-6269 (2019). https://doi.org/10.1021/acs.nanolett.9b02369
5. F. Paolucci, G. De Simoni, P. Solinas, E. Strambini, C. Puglia, N. Ligato, F. Giazotto, "Field-effect control of metallic superconducting systems", *AVS Quantum Science* **1**, 016501 (2019). https://doi.org/10.1116/1.5129364
6. C. Puglia, G. De Simoni, F. Giazotto, "Electrostatic control of phase slips in Ti Josephson nanotransistors", *Phys. Rev. Appl*. **13**, 054026 (2020). https://doi.org/10.1103/PhysRevApplied.13.054026
7. G. De Simoni, C. Puglia, F. Giazotto, "Niobium Dayem nano-bridge Josephson gate-controlled transistors", *Appl. Phys. Lett*. **116**, 242601 (2020). https://doi.org/10.1063/5.0011304
8. C. Puglia, G. De Simoni, N. Ligato, F. Giazotto, "Vanadium gate-controlled Josephson half-wave nanorectifier", *Appl. Phys. Lett*. **116**, 252601 (2020). https://doi.org/10.1063/5.0013512
9. L. Bours, M. T. Mercaldo, M. Cuoco, E. Strambini, F. Giazotto "Unveiling mechanisms of electric field effects on superconductors by a magnetic field response", *Phys. Rev. Res*. **2**, 033353 (2020). https://doi.org/10.1103/PhysRevResearch.2.033353
10. M. Rocci, G. De Simoni, C. Puglia, D. Degli Esposti, E. Strambini, V. Zannier, L. Sorba, F. Giazotto "Gate-controlled suspended titanium nanobridge supercurrent transistor", *ACS Nano* **14**, 12621-12628 (2020). https://doi.org/10.1021/acsnano.0c05355
11. M. Rocci, D. Suri, A. Kamra, G. Vilela, Y. Takamura, N. M. Nemes, J. L. Martinez, M. G. Hernandez, J. S. Moodera, "Large enhancement of critical current in superconducting devices by gate voltage", *Nano Lett*. **21**, 216-221 (2020). https://dx.doi.org/10.1021/acs.nanolett.0c03547
12. L. D. Alegria, C. G. L. Bøttcher, A. K. Saydjari, A. T. Pierce, S. H. Lee, S. P. Harvey, U. Vool, A. Yacoby, "High-energy quasiparticles injection into mesoscopic superconductors", *Nat. Nanotech*. **16**, 404-408 (2021). https://doi.org/10.1038/s41565-020-00834-8
13. M. F. Ritter, A. Fuhrer, D. Z. Haxell, S. Hart, P. Gumann, H. Riel, F. Nichele, "A superconducting switch actuated by injection of high-energy electrons", *Nat. Commun*. **12**, 1266 (2021). https://doi.org/10.1038/s41467-021-21231-2
14. I. Golokolenov, A. Guthrie, S. Kafanov, Y. A. Pashkin, V. Tsepelin, "On the origin of the controversial electrostatic field effect in superconductors", *Nat. Commun*. **12**, 2747 (2021). https://doi.org/10.1038/s41467-021-22998-0
15. P. Orús, V. M. Fomin, J. M. De Teresa, R. Córdoba, "Critical current modulation induced by an electric field in superconducting tungsten-carbon nanowires", *Sci. Rep*. **11**, 17698 (2021). https://doi.org/10.1038/s41598-021-97075-z
16. G. De Simoni, S. Battisti, N. Ligato, M. T. Mercaldo, M. Cuoco, F. Giazotto, "Gate control of the current-flux relation of a Josephson quantum interferometer based on proximitized metallic nanojunctions", *ACS Appl. Electron. Mater*. **3**, 3927-3935 (2021). https://doi.org/10.1021/acsaelm.1c00508
17. T. Elalaily, O. Kürtössy, Z. Scherübl, M. Berke, G. Fülöp, I. E. Lukács, T. Kanne, J. Nygård, K. Watanabe, T. Tanighuchi, P. Makk, S. Csonka, "Gate-controlled supercurrent in epitaxial Al/InAs nanowires", *Nano Lett*. **21**, 9684-9690 (2021). https://doi.org/10.1021/acs.nanolett.1c03493
18. F. Paolucci, F. Crisa, G. De Simoni, L. Bours, C. Puglia, E. Strambini, S. Roddaro, F. Giazotto, "Electrostatic field-driven supercurrent suppression in ionic-gated metallic superconducting nanotransistors", *Nano Lett*. **21**, 10309-10314 (2021). https://doi.org/10.1021/acs.nanolett.1c03481
19. J. Basset, O. Stanisavljevic, M. Kuzmanovic, J. Gabelli, C. H. L. Quay, J. Esteve, M. Aprili, "Gate-assisted phase fluctuations in all-metallic Josephson junctions", *Phys. Rev. Res*. **3**, 043169 (2021). https://doi.org/10.1103/PhysRevResearch.3.043169

20. M. F. Ritter, N. Crescini, D. Z. Haxell, M. Hinderling, H. Riel, C. Bruder, A. Fuhrer, F. Nichele, "Out-of-equilibrium phonons in gated superconducting switches", *Nat. Electron*. **5**, 71-77 (2022). https://doi.org/10.1038/s41928-022-00721-1
21. G. Catto, W. Liu, S. Kundu, V. Lahtinen, V. Vesterinen, M. Mottonen, "Microwave response of a metallic superconductor subject to a high-voltage gate electrode", *Sci. Rep*. **12**, 6822 (2022). https://doi.org/10.1038/s41598-022-10833-5
22. T. Elalaily, M. Berke, M. Kedves, G. Fülöp, Z. Scherübl, T. Kanne, J. Nygård, P. Makk, S. Csonka, "Signatures of gate-driven out of equilibrium superconductivity in Ta/InAs nanowires", *ACS Nano* **17**, 5528 (2023). https://doi.org/10.1021/acsnano.2c10877
23. T. Jalabert, E. F. C. Driessen, F. Gustavo, J. L. Thomassin, F. Levy-Bertrand, C. Chapelier, "Thermalization and dynamic of high-energy quasiparticles in a superconducting nanowire", *Nat. Phys*. **19**, 956-960 (2023). https://doi.org/10.1038/s41567-023-01999-4
24. H. Du, Z. Xu, Z. Wei, D. Li, S. Chen, W. Tian, P. Zhang, Y.-Y. Lyu, H. Sun, Y.-L. Wang, H. Wang, and P. Wu, "High-energy electron local injection in top-gated metallic superconductor switch", *Supercond. Sci. Technol*. **36**, 095005 (2023). https://doi.org/10.1088/1361-6668/ace65f
25. S. Yu, L. Chen, Y. Pan, Y. Wang, D. Zhang, G. Wu, X. Fan, X. Liu, L. Wu, L. Zhang, W. Peng, J. Ren, Z. Wang, "Gate-tunable critical current of the three-dimensional niobium nanobridge Josephson junction", *Nano Lett*. **23**, 8043 (2023). https://doi.org/10.1021/acs.nanolett.3c02015
26. L. Ruf, T. Elalaily, C. Puglia, Yu. P. Ivanov, F. Joint, M. Berke, A. Iorio, P. Makk, G. De Simoni, S. Gasparinetti, G. Divitini, S. Csonka, F. Giazotto, E. Scheer, A. Di Bernardo, "Effects of fabrication routes and material parameters on the control of superconducting currents by gate voltage", *APL Mater*. **11**, 091113 (2023). https://doi.org/10.1063/5.0159750
27. J. Koch, C. Cirillo, S. Battisti, L. Ruf, Z. Makhdoumi Kakhaki, A. Paghi, A. Gulian, S. Teknowijoyo, G. De Simoni, F. Giazotto, C. Attanasio, E. Scheer, A. Di Bernardo, "Gate-controlled supercurrent effect in dry-etched Dayem bridges of non-centrosymmetric niobium rhenium", Accepted in *Nano Reearch* (2024), https://doi.org/10.1007/s12274-024-6576-7.
28. T. Elalaily, M. Berke, I. Lilja, A. Savin, G. Fülöp, L. Kupás, T. Kanne, J. Nygård, P. Makk. P. Hakonen, S. Csonka, "Switching dynamics in Al/InAs nanowire-based gate-controlled superconducting transistor" (2023), pre-print available at https://arxiv.org/abs/2312.15453
29. L. Zhang, H. Li, J. Tan, Y. Guan, Q. Chen, H. Wang, B. Zhang, X. Tu, Q.-Y. Zhao, X.-Q. Jia, L. Kang, Z. Yang, J. Chen, P.-H. Wu, "An incoherent superconducting nanowire photon detector revealing the controversial gate-controlled superconductivity" (2023), pre-print available at https://www.researchsquare.com/article/rs-3718133/v1
30. M. T. Mercaldo, P. Solinas, F. Giazotto, M. Cuoco, "Electrically tunable superconductivity through surface orbital polarization", *Phys. Rev. Appl*. **14**, 034041 (2020). https://doi.org/10.1103/PhysRevApplied.14.034041
31. P. Solinas, A. Amoretti, F. Giazotto, "Sauter-Schwinger effect in a Bardeen-Cooper-Schrieffer superconductor", *Phys. Rev. Lett.* **126**, 117001 (2021). https://doi.org/10.1103/PhysRevLett.126.117001
32. L. Chirolli, T. Cea, F. Giazotto, "Impact of electrostatic fields in layered crystalline BCS superconductors", *Phys. Rev. Res*. **3**, 023135 (2021). https://doi.org/10.1103/PhysRevResearch.3.023135
33. M. T. Mercaldo, F. Giazotto, M. Cuoco, "Spectroscopic signatures of gate-controlled superconducting phases", *Phys. Rev. Res*. **3**, 043042 (2021). https://doi.org/10.1103/PhysRevResearch.3.043042
34. A. Amoretti, D. K. Brattan, N. Magnoli, L. Martinoia, I. Matthaiakakis, P. Solinas, "Destroying superconductivity in thin films with an electric field", *Phys. Rev. Res*. **4**, 033211 (2022). https://doi.org/10.1103/PhysRevResearch.4.033211
35. S. Chakraborty, D. Nikolić, J. C. Cuevas, F. Giazotto, A. Di Bernardo, E. Scheer, M. Cuoco, W. Belzig, "Microscopic theory of supercurrent suppression by gate-controlled surface depairing", *Phys. Rev. B* **108**, 184508 (2023).
36. A. Zaccone, V. M. Fomin, "Theory of superconductivity in thin films under an external electric field" (2023), pre-print available at https://arxiv.org/abs/2312.13059.
37. O. A. Mukhanov, "Energy-efficient single flux quantum technology", *IEEE Trans. Appl. Supercond*. **21**, 760-769 (2011). https://doi.org/10.1109%2FTASC.2010.2096792

38. V. K. Semenov, Y. A. Polyakov, S. K. Tolpygo, "New AC-powered SFQ digital circuits", *IEEE Trans. Appl. Supercond.* **25**, 1301507 (2015). https://doi.org/10.1109/TASC.2014.2382665
39. W. Chen, A. V. Rylyakov, V. Patel, J. E. Lukens, K. K Likharev, "Rapid single flux quantum T-flip flop operating up to 770 GHz", *IEEE Trans. Appl. Supercond*. **9**, 3212-3215 (1999). https://doi.org/10.1109/77.783712
40. L. Ruf, C. Puglia, G. De Simoni, Yu P. Ivanov, T. Elalaily, F. Joint, M. Berke, J. Koch, A. Iorio, S. Khorshidian, P. Makk, S. Gasparinetti, S. Csonka, W. Belzig, M. Cuoco, G. Divitini, F. Giazotto, E. Scheer, A. Di Bernardo, "Gate-control of superconducting current: relevant parameters and perspectives", pre-print available at https://arxiv.org/abs/2302.13734
41. V. Kursun, E. G. Friedman in *Multi-voltage CMOS circuit design*, Wiley, New York (2006). https://doi.org/10.1002/0470033371
42. F. B. Hagedorn, P. M. Hall. "Right-angle bends in thin strip conductors", *J. Appl. Phys.* **34**, 128-133 (1963). https://doi.org/10.1063/1.1729052
43. J. R. Clem, K. K. Berggren, "Geometry-dependent critical current in superconducting nanocircuits", *Phys. Rev. B* **84**, 174510 (2011). http://dx.doi.org/10.1103/PhysRevB.84.174510
44. J. Y. Gu, J. A. Caballero, R. D. Slater, R. Loloee, W. P. Pratt, Jr, "Direct measurement of quasiparticle evanescent waves in a dirty superconductor", *Phys. Rev. B* **66**, 140507 (R) (2002). https://doi.org/10.1103/PhysRevB.66.140507
45. M. Kimura, T. Ohmi, "Conduction mechanism and origin and stress-induced leakage current in thin silicon dioxide films", J. Appl. Phys. 80, 6360 (1996). https://doi.org/10.1063/1.363655
46. T. Ishida, N. Tega, Y. Mori, H. Miki, T. Mine, H. Kume, K. Torii, R.Yamada, K. Shiraishi, "Mechanism of state transition of a defect causing random-telegraph-noise-induced fluctuation in stress-induced leakage current of $SiO_2$ films", *Jpn. J. Appl. Phys,* **53**, 08LB01 (2014). https://www.doi.org/10.7567/JJAP.53.08LB01
47. Y. Ma, J. Bi, H. Wang, L. Fan, B. Zhao, L. Shen, M. Liu, "Mechanism of random telegraph noise in 22-nm FDSOI-based MOSFET at cryogenic temperatures", *Nanomat.,* **12**, 4344 (2022). https://doi.org/10.3390/nano12234344
48. R. Yamada, T. J. King, "Variable stress-induced leakage current and analysis of anomalous charge loss for flash memory application", *IEEE Int. Rel. Phys. Symp. Proc. 2003*, Dallas, TX, 491-496 (2003). https://doi.org/10.1109/RELPHY.2003.1197797
49. T. Ishida, N. Tega, Y. Mori, H. Miki, T. Mine, H. Kume, K. Torri, m. Muraguchi, Y. Takada, K. Shiraishi, R. Yamada*,* "A new insight into the dynamic fluctuation mechanism of stress-induced leakage current", *IEEE Int. Rel. Phys. Symp. 2008*, Phoenix, AZ, 604-609 (2008). https://www.doi.org/10.1109/RELPHY.2008.4558953

Supplementary Information for

# High-performance gate-controlled superconducting switches: large output voltage and reproducibility

L. Ruf[1], E. Scheer[1,†], A. Di Bernardo[1,2,*]

1. *Department of Physics, University of Konstanz, Universitätsstraße 10, 78464 Konstanz, Germany.*
2. *Dipartimento di Fisica "E. R. Caianiello", Università degli Studi di Salerno, via Giovanni Paolo II 132, 84084 Fisciano (SA), Italy.*

[†]Email: elke.scheer@uni-konstanz.de
[*]Email: angelo.dibernardo@uni-konstanz.de

**This file contains:**
- Supplementary Table S1 with the main parameters of all the devices investigated.
- Supplementary Figs. S1 to S9

| Dev id | $R_N$ (Ω) | $T$ (K) | $T_m$ (K) | $I_{c0}$ (μA) | $I_{r0}$ (μA) | $V^*_{G,onset}$ (V) | $V^*_{G,offset}$ (V) | $I_{leak}$ @ $V^*_{G,onset}$ (nA) | $I_{leak}$ @ $V^*_{G,offset}$ (nA) | Gate sep. (nm) | Width $w_S$ (nm) |
|---|---|---|---|---|---|---|---|---|---|---|---|
| D1 | 475.9 | 2.24 | 1.39 | 12.2 | 2.30 | 0.85 | 1.60 | 1.4 | 14.6 | ~50 | 190 |
| D2 | 202.0 | 4.08 | 1.64 | 20.4 | 16.46 | 15.6 | 19.4 | 0.71 | 4.5 | 70 | 190 |
| D3 | 203.1 | 3.68 | 3.09 | 15.7 | 8.73 | 24.0 | 32.5 | 0.04 | n/a | < 100 | ~450 |
| D4 | 164.1 | 5.08 | 3.10 | 187.0 | 28.00 | 24.3 | 32.5 | 0.17 | 1.6 | ~50 | 450 |
| D5 | 134.6 | 5.14 | 3.14 | 273.0 | 30.07 | 27.5 | 33.6 | 0.21 | 1.3 | ~50 | ~450 |
| D6 | 144.0 | 5.17 | 3.10 | 280.0 | 28.00 | 14.5 | 17.1 | 0.27 | 2.2 | ~50 | ~450 |
| D7 | 142.5 | 5.14 | 3.10 | 256.5 | 30.07 | 26.6 | 34.1 | 0.15 | 1.6 | ~50 | ~450 |
| D8 | 132.6 | 5.22 | 3.10 | 300.0 | 32.15 | 25.1 | 30.1 | 0.21 | 1.6 | ~50 | ~450 |
| D9 | 127.9 | 5.34 | 3.16 | 326.2 | 31.5 | 20.6 | 27.1 | 0.14 | 1.5 | ~50 | ~450 |
| D10 | 78.6 | 6.34 | 3.10 | 1245 | 94.00 | 21.2 | 29.7 | 0.22 | 4.0 | ~50 | 550 w/FIB cut of 90 |
| D11 | 61.5 | 7.12 | 3.6 | 2570 | 136.0 | 5.7 | 9.5 | 0.2 | 25.3 | ~50 | 550 |
| D12 | 61.2 | 7.57 | 4.0 | 2300 | 136.0 | 9.7 | 13.5 | 0.6 | 18 | ~50 | 550 |
| D13 | 123.4 | 5.03 | 3.14 | 272 | 33.0 | 28.5* | 37* | 0.07* | 0.9* | ~50 | 550 |

**Table S1 – Parameters of the gate-controlled Nb devices investigated in this study.** For each device, with identification number specified in the first column, the table reports the normal-state resistance $R_N$ measured at 10 K, the superconducting critical temperature $T_c$ defined as the temperature at which $R_N$ drops by 90%, the temperature at which the device has been characterized for the GCS ($T_m$), the critical current ($I_{c0}$) and retrapping current ($I_{r0}$) measured at $T_m$ without gate voltage applied, the actual gate voltage applied (i.e., corrected for the voltage drop over the wiring resistance; see main text) needed to reduce $I_{c0}$ by 10% ($V^*_{G,onset}$) and by 90% ($V^*_{G,offset}$), the leakage current $I_{leak}$ values measured at $V^*_{G,onset}$ and $V^*_{G,onset}$, the separation between gate and S constriction, and the width ($w_S$) of the S constriction. The devices reported in the first two rows have been measured in a cryostat manufactured by Cryogenic, whilst all the other devices have been measured in another cryostat manufactured by ICE Oxford (see also Methods in the main text). The symbol '*' refers to parameter values measured after $I_{leak}$-induced training of the device (see main text).

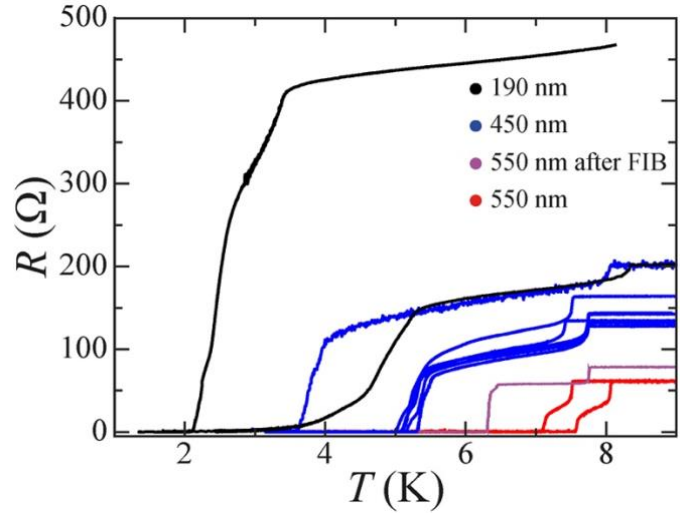

**Figure S1 – Low-temperature transport data of all devices**. Resistance versus temperature $R(T)$ curves measured for all Nb Dayem bridges fabricated in this study and listed in Table S1, with width $w_s$ = 190 nm (black; 2 devices), 450 nm (blue; 7 devices); 550 nm without (red; 2 devices) and with 90-nm-wide cut made by focused ion beam on the Nb constriction opposite from the gate (purple; 1 device).

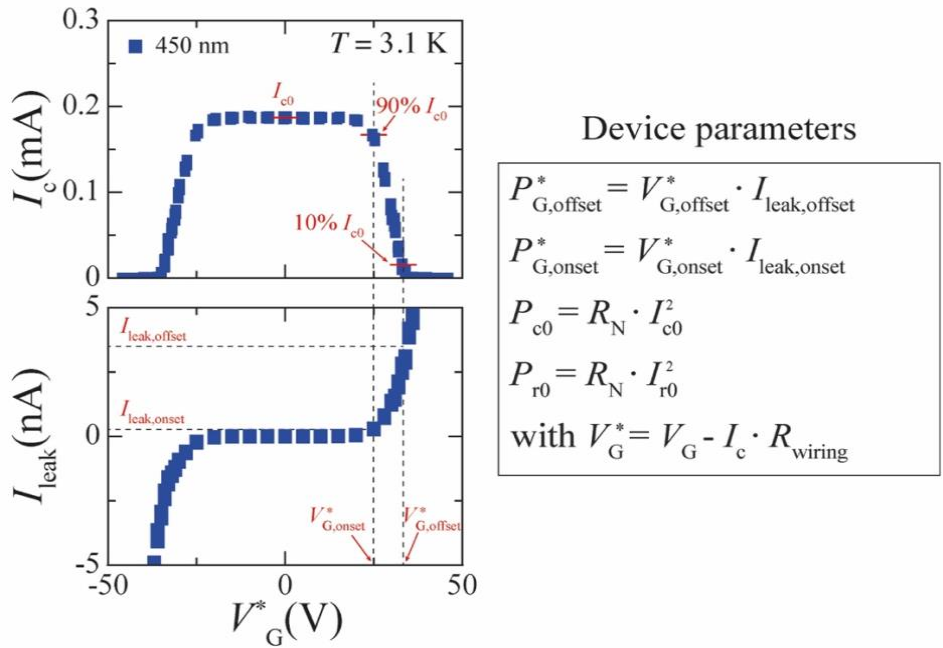

**Figure S2 – Definition of parameters of a GCS device**. Critical current versus effective applied gate voltage, $I_c(V_G^*)$, and leakage current versus $V_G^*$, $I_{leak}(V_G^*)$, measured for a GCS device with width $w_S$ = 450 nm at $T$ = 3.1 K. From the $I_c(V_G^*)$ curve, the $V_G^*$ values at which the $I_c$ at $V_G^*$ = 0 ($I_{c0}$) is reduced by 10% ($V_{G,onset}^*$) and by 90% ($V_{G,offset}^*$) are identified, as shown by the dashed lines in the graph. The values of $I_{leak}$ at $V_{G,onset}^*$ ($I_{leak,onset}$) and $V_{G,offset}^*$ ($I_{leak,offset}$) are then also derived from the $I_{leak}(V_G^*)$ characteristic. The box on the right reports a list of the main device parameters that can be calculated once $I_{c0}$, $V_{G,onset}^*$, $V_{G,offset}^*$, $I_{leak,onset}$ and $I_{leak,offset}$ have been obtained. The meaning of these parameters and of the other variables listed is explained in the main text.

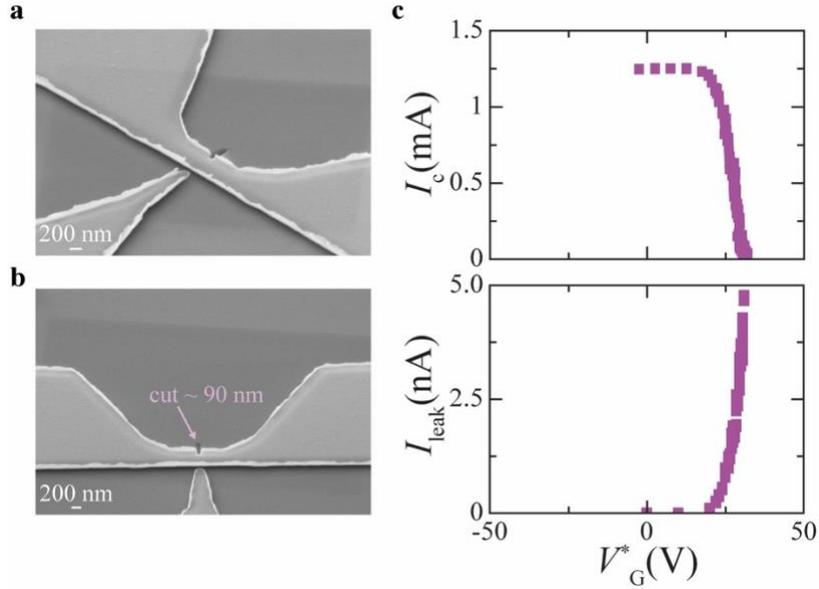

**Figure S3 – Characterization of a FIB-cut device**. (a-b) Scanning electron micrograph images of a Nb bridge with original width $w_s$ = 550 nm reduced by ~ 90 nm using a focused ion beam with $Ga^+$ ions. (c) Critical current versus gate voltage $I_c(V^*_G)$ (top panel) and leakage current versus gate voltage $I_c(V^*_G)$ measured on the same device at a temperature $T$ ~ 3.1 K.

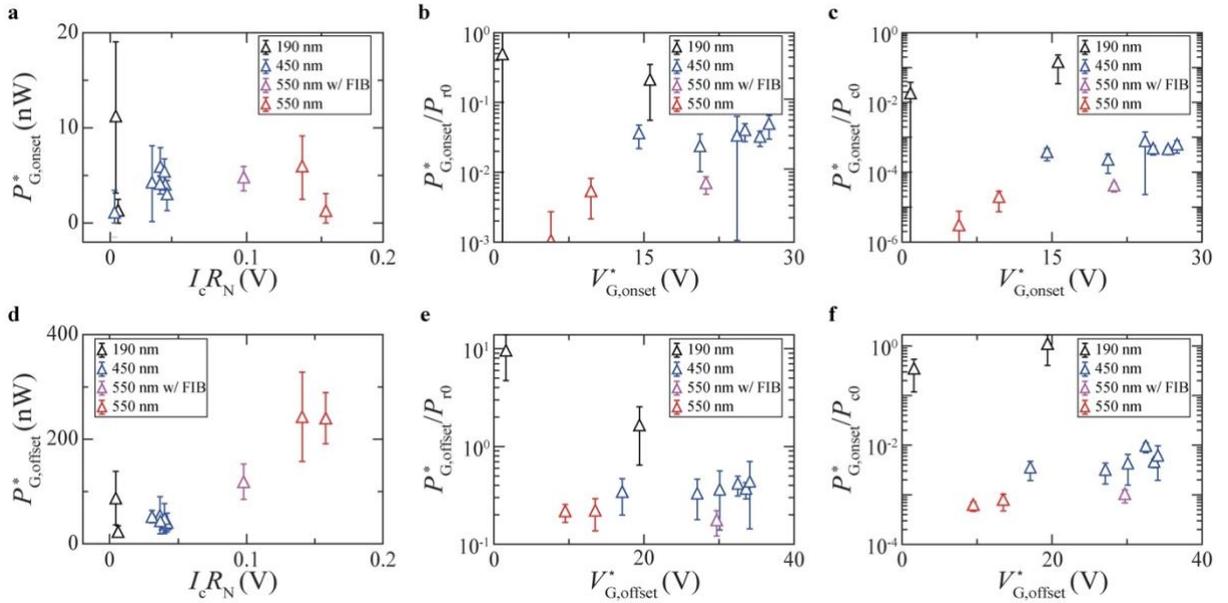

**Figure S4 – Performance parameters of all Nb devices studied.** (a-c) Onset power dissipated by the gate $P^*_{G,onset}$ as a function of the characteristic voltage $I_cR_N$ (a), of the effective gate voltage $V^*_{G,onset}$ for 10% $I_c$ suppression (b), and of the effective $V^*_{G,offset}$ for 90% $I_c$ suppression (c) measured for all devices investigated with different width $w_S$ (as specified in the legend) and at $T$ ~ 3.1 K. (d-f) Dependence of the offset power dissipated by the gate $P^*_{G,offset}$ on the same parameters $I_cR_N$ (d), $V^*_{G,onset}$ (e), $V^*_{G,onset}$ (f) for the same devices reported in (a-d) (see legends for $w_S$ values) and measured at the same $T$ ~ 3.1 K.

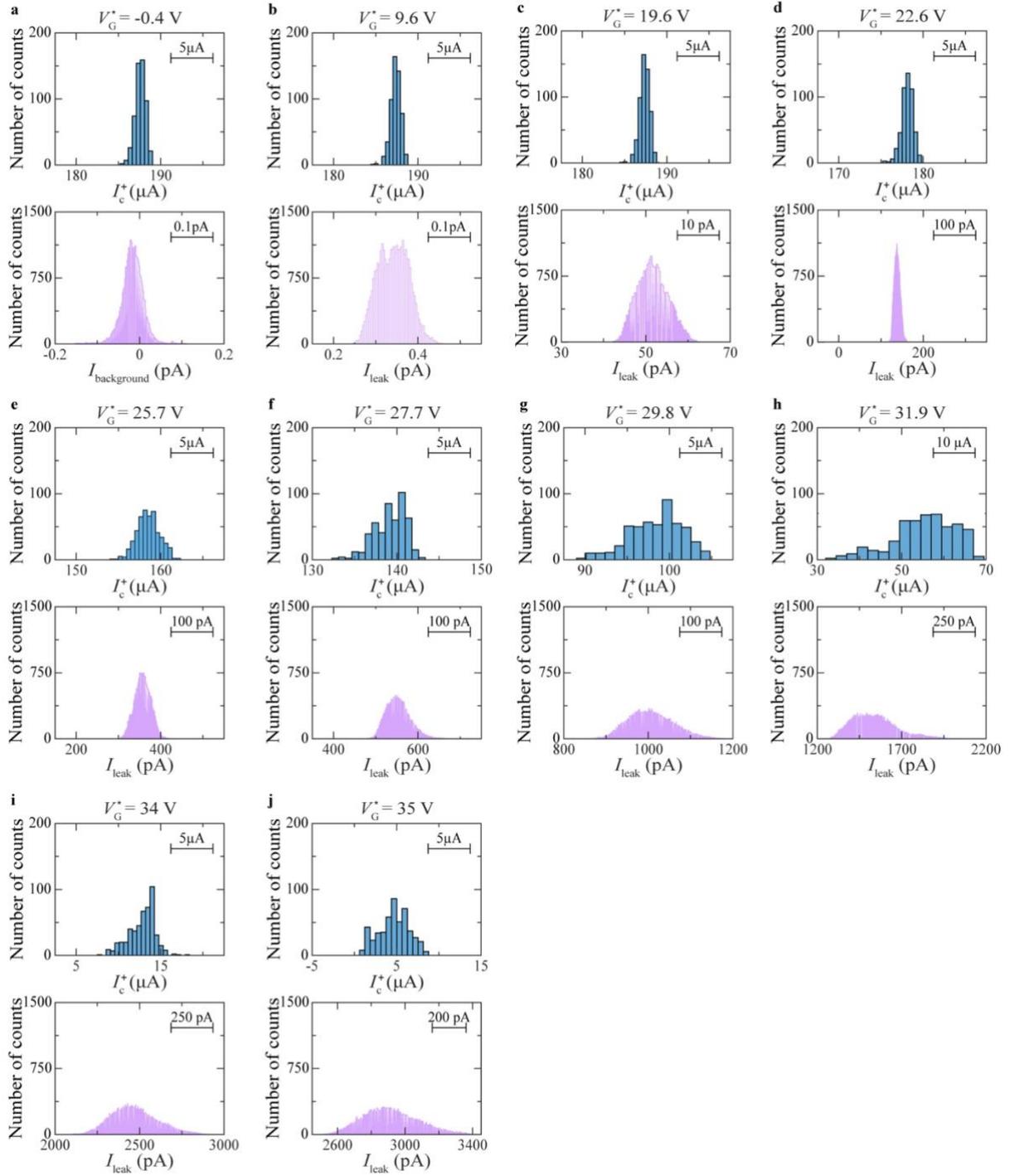

**Figure S5 – Gate dependence of switching current and leakage current distributions.** (a-j) Distributions of the positive critical current $I_c^+$ (top part of the panel) and of the leakage current $I_{leak}$ (bottom part of the panel) measured for the same device shown in Figs. S7 and S8 at different gate voltage $V_G^*$ ($V_G^*$ values reported on top of each panel) and temperature $T \sim 3.1$ K. $I_c^+$ has been measured whilst upsweeping the bias current from 0 to positive values.

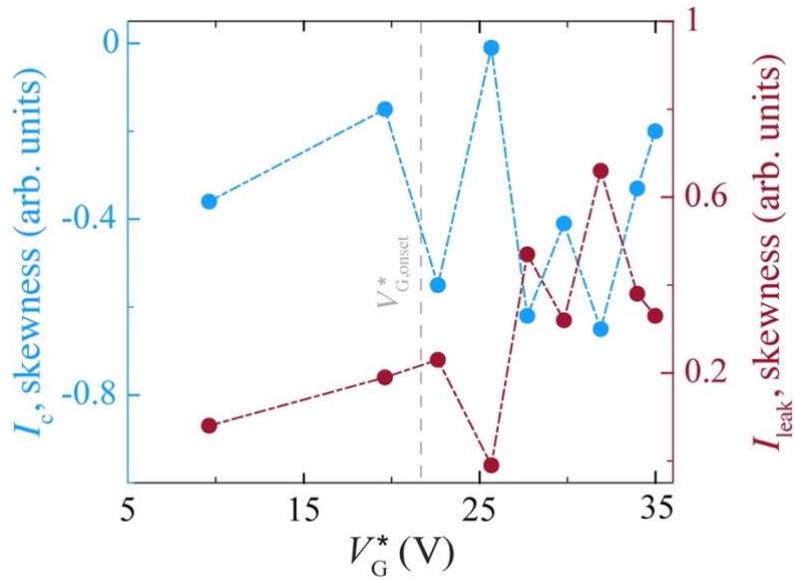

**Figure S6 – Anticorrelation between leakage and critical current distributions.** Dependence of the skewness of the distributions of the critical current $I_c$ (left axis; blue curve) and of the skewness of the leakage current $I_{leak}$ (right axis; red curve) measured for a gated Nb device as a function of the effective gate voltage $V_G^*$ applied. The two curves show a clear anticorrelation concomitant with the occurrence of the GCS for $V_G^* > V_{G,onset}^*$ (dashed gray line).

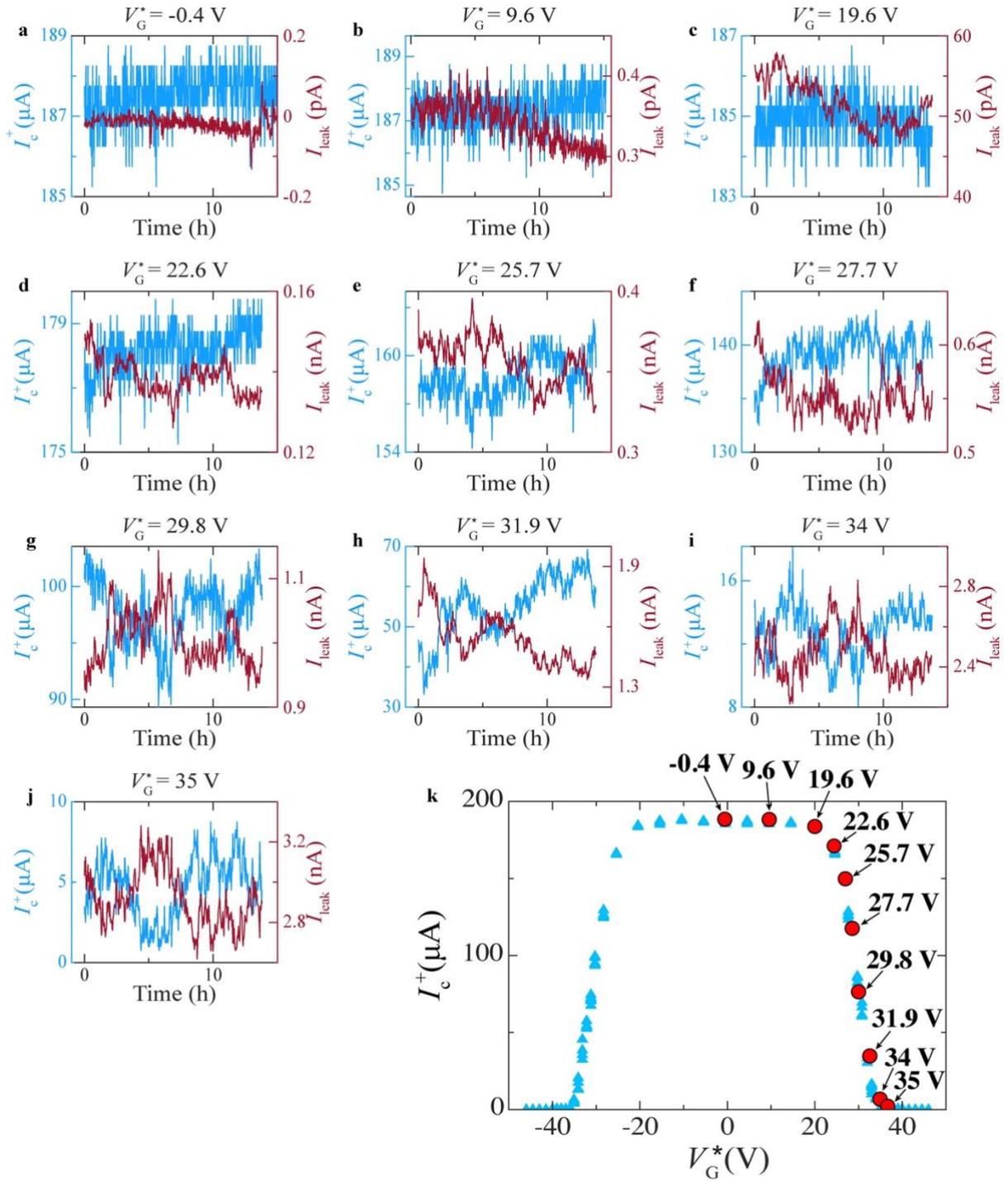

**Figure S7 – Time evolution of $I_c^+$ and $I_{leak}$.** (a-j) Positive critical current $I_c^+$ and leakage current $I_{leak}$ measured as a function of time at fixed applied gate voltage $V_G^*$ ($V_G^*$ values indicated on top of each panel) and temperature $T \sim 3.1$ K. At a given time $t$, $I_c^+$ is obtained from the current versus voltage $I(V)$ characteristics measured by upsweeping the bias current $I$ from 0 A to positive values. (k) $I_c^+(V_G^*)$ curve showing the observation of a GCS and reporting the $V_G^*$ values (marked with red circles) at which the measurement data in panels from (a) to (j) have been obtained in the same device.

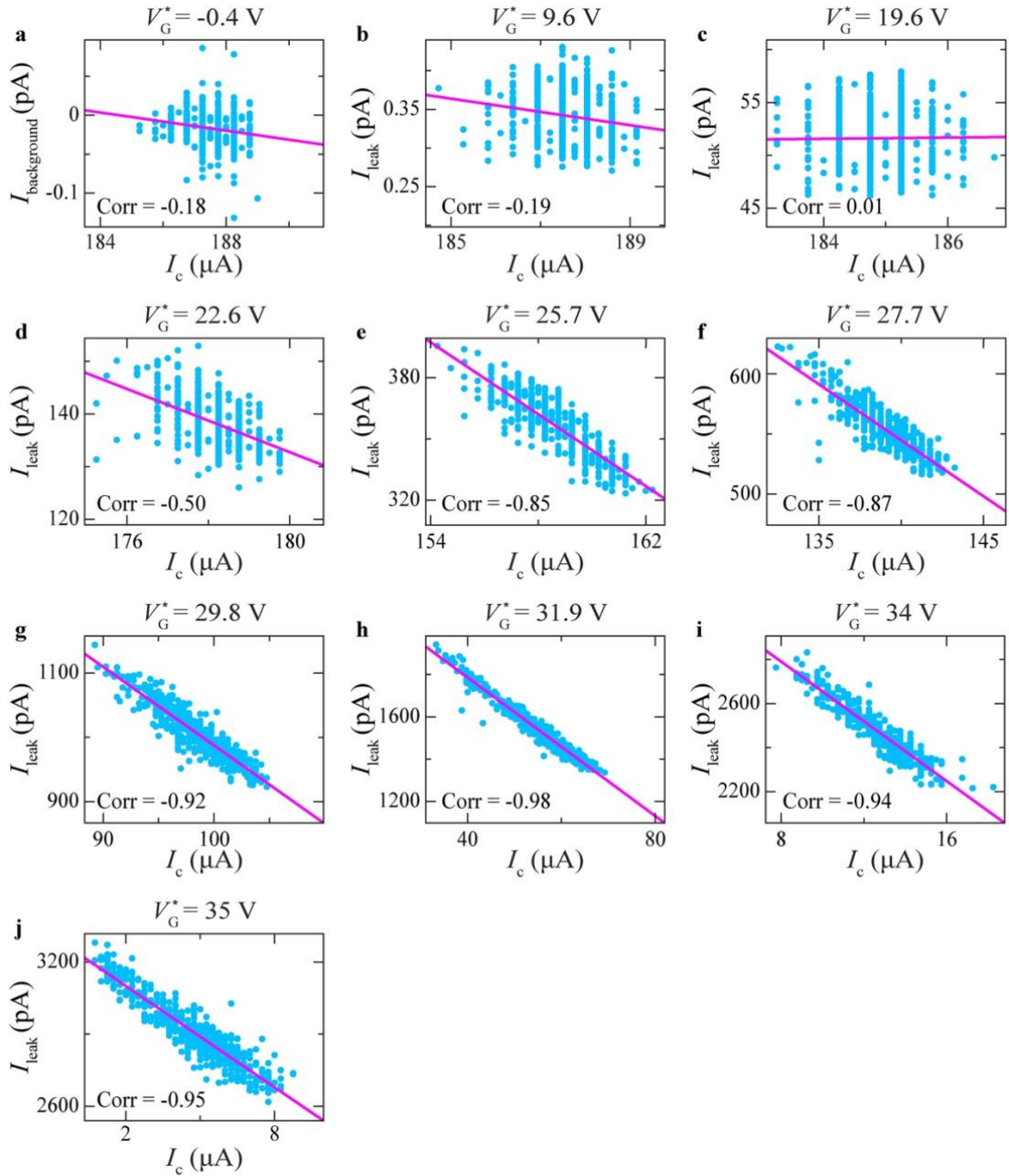

**Figure S8 – Anticorrelation between $I_c$ and $I_{leak,sweep}$.** (a-j) Average leakage current per sweep $I_{leak,sweep}$ as a function of the positive critical current $I_c^+$ measured for the same device shown in Figs. S5 and S7 at different gate voltage $V_G^*$ ($V_G^*$ values reported on top of each panel) and temperature $T \sim 3.1$ K. For each panel, the applied $V_G^*$ is the same as that of the panel labelled with the same letter in Figs. S5 and S7, and the correlation factor obtained is reported in the bottom-left corner. For panel (a), the noise on $I_{leak}$ due to the instrumental setup at $V_G^* = 0$ is reported.

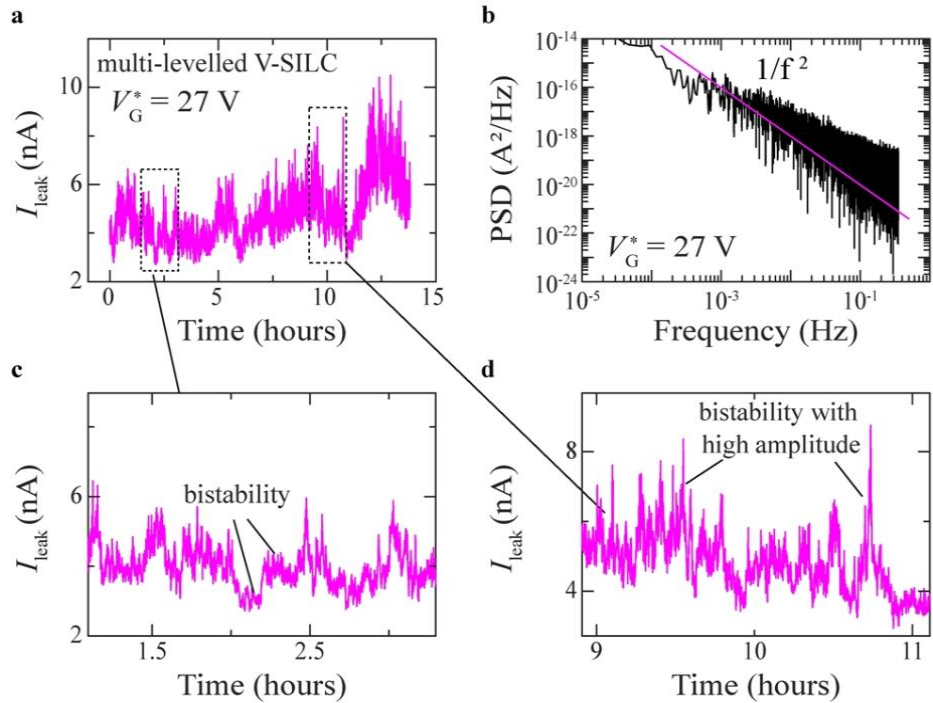

**Figure S9 – Instabilities in leakage current and relation to SiO$_2$/Si substrate.** (a) Leakage current measured over time $I_{leak}$ for a Nb device under an applied gate voltage $V_G^* = 27$ V above the onset voltage of the GCS for the device. (b) Power spectral density (PDS) corresponding to the signal in (a) showing trend close to $1/f^2$-type noise. (c-d) $I_{leak}$ signal in (a) over selected time periods (indicated by dashed boxes) showing multiple instabilities consistent observed for multi-levelled variable stress-induced leakage current (V-SILC) in SiO$_2$, in addition to bistabilities with small (c) or large (d) amplitudes.